\documentstyle[12pt,world_sci]{article}
\input epsf
\begin{document}
\title{OVERVIEW OF THE STANDARD MODEL}
\author{JONATHAN L. ROSNER \\
{\em Enrico Fermi Institute and Department of Physics,
University of Chicago\\
5640 S. Ellis Ave., Chicago, IL 60637, USA}}
\maketitle
\setlength{\baselineskip}{2.6ex}
\def \g{{\rm GeV}}
\vspace{-1.9in}
\rightline{EFI 94-59}
\rightline{hep-ph/9411396}
\rightline{November 1994}
\vspace{-0.667in}
\leftline{Presented at Joint US - Polish Workshop}
\leftline{~~~on Physics from Planck Scale to Electroweak Scale}
\leftline{Warsaw, Poland, 21 -- 24 September, 1994}
\leftline{Proceedings to be published by World Scientific}
\vspace{1.4in}

\begin{abstract}
\small
A brief overview is presented of the standard model of electroweak
interactions, including precision tests, the role of the
Cabibbo-Kobayashi-Maskawa (CKM) matrix in describing CP violation, windows on
intermediate-scale physics, and the nature of the Higgs phenomenon.
\end{abstract}

\section{Introduction}

The standard model of electroweak interactions has guided experimental
discoveries for more than ten years, starting with the observation of the $W$
and $Z$ bosons and culminating this past year with evidence for the top quark.
The new particles' properties have been very close to those anticipated by
theory:  reassuring to some physicists, frustrating to others. In this brief
overview we describe the present state of electroweak theory and indicate ways
in which new insights could emerge from experiments within the next decade.

Precise electroweak tests, described in Sec.~2, show promise of shedding light
on the Higgs, and restrict our ability to add large numbers of new particles to
the theory, as in some models of dynamical electroweak symmetry breaking. A
plausible theory of CP violation, based on phases in the
Cabibbo-Kobayashi-Maskawa (CKM) matrix, can be checked by many forthcoming
experiments on rare $K$ decays and properties of $B$ mesons, as noted in
Sec.~3.

While electroweak physics is based on physics below the TeV scale, unification
of electroweak and strong interactions must take place at scales above
$10^{15}$ GeV, or protons would decay too rapidly.  Although many schemes
envision a ``desert'' between 1 TeV and the unification scale, we describe in
Sec.~4 several probes of intermediate scales (such as the interesting range
$10^9 - 10^{12}$ GeV), including the study of neutrino masses, the
investigation of baryogenesis in the early Universe, and the search for light
pseudoscalar particles (``axions'').

The mechanism of electroweak symmetry breaking is still unknown.  Does it
proceed via fundamental scalars in the theory, whose masses are protected from
large radiative corrections by some mechanism such as supersymmetry, or is
it dynamically generated through new interactions which are strong at the
TeV scale?  These questions are briefly considered in Sec.~5.  We summarize
in Sec.~6.  More details on some of the topics presented here may be found in
Ref.~1.

\section{Precise electroweak tests}

\subsection{In celebration of the top quark}

In April of this past year the CDF Collaboration at Fermilab presented
evidence\cite{CDFtop} for a top quark with mass $m_t = 174 \pm 10 ^{+13}_{-12}$
GeV.  The measured cross section is slightly higher than (though consistent
with) theoretical expectations based on quantum chromodynamics (QCD).
Mechanisms which could give an elevated cross section include the production of
more than one flavor of new quark and schemes involving dynamical symmetry
breaking\cite{EL,HP}.  The mass lies in the middle of the range anticipated on
the basis of electroweak radiative corrections (to be discussed in this
section) and analyses of the CKM matrix (to be discussed in the next).  But why
is the top quark so heavy?

\subsection{Quark and lepton masses}

\begin{figure}
\centerline{\epsfysize = 4in \epsffile{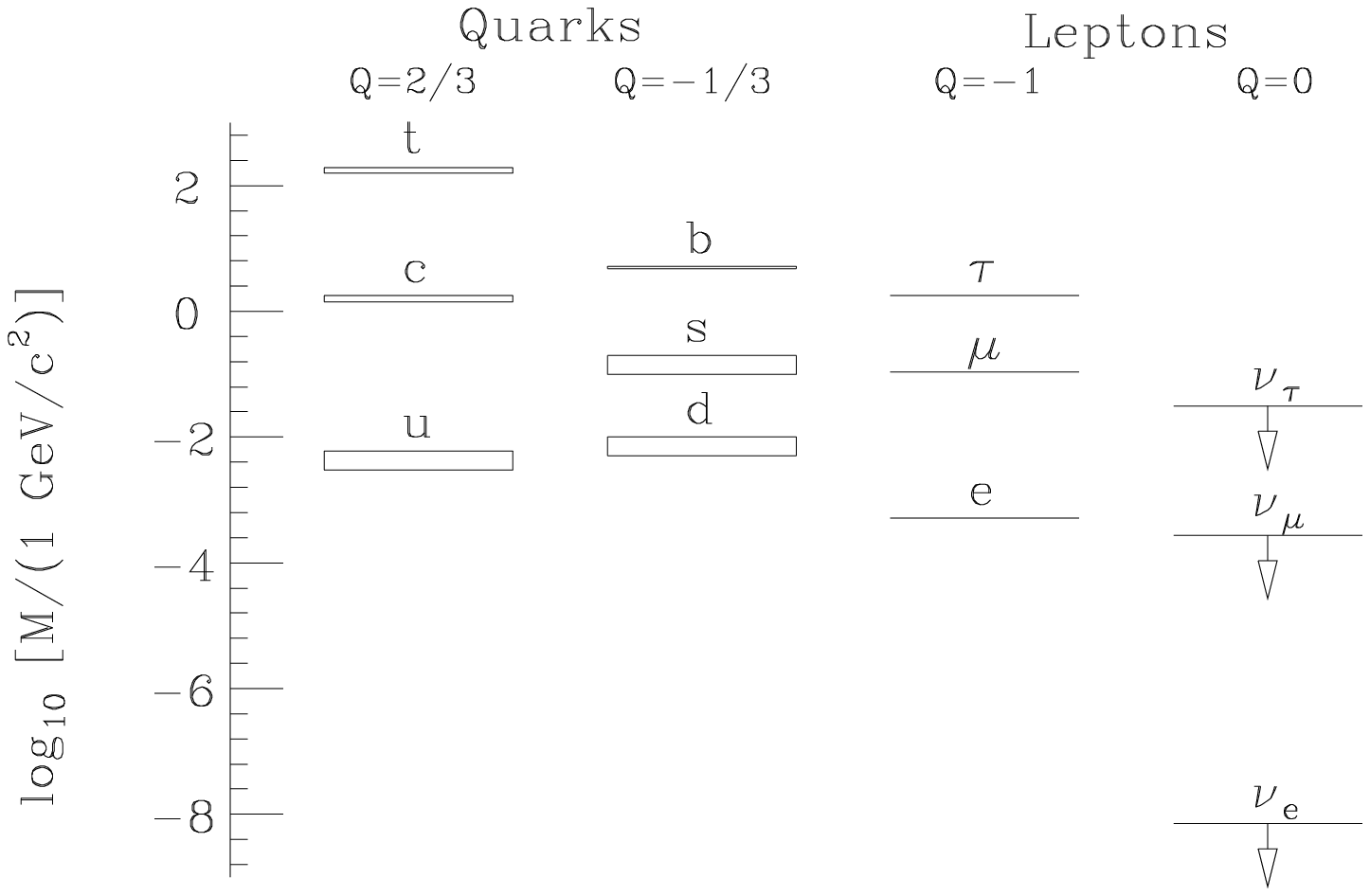}}
\caption{Masses of quarks and leptons on a logarithmic scale.  Widths of bars
denote uncertainties in quark masses.}
\end{figure}

The top quark is the last quark to fit into a set of three families of quarks
and leptons, whose masses are shown in Fig.~1:
\begin{equation}
\left ( \begin{array}{c} u \\ d \end{array} \right) ~~;~~~
\left ( \begin{array}{c} c \\ s \end{array} \right) ~~;~~~
\left ( \begin{array}{c} t \\ b \end{array} \right) ~~;
\end{equation}
\begin{equation}
\left ( \begin{array}{c} \nu_e \\ e \end{array} \right) ~~;~~~
\left ( \begin{array}{c} \nu_\mu \\ \mu \end{array} \right) ~~;~~~
\left ( \begin{array}{c} \nu_\tau \\ \tau \end{array} \right) ~~.
\end{equation}

Only the $\nu_\tau$ has not yet been directly observed.  The fractional error
on the top mass is less than that for the light quarks $u,~d$, and $s$, whose
masses are obscured by QCD effects.  In logarithmic terms, the $t$ mass seems
not all that anomalous, but we have no theory for the pattern in Fig.~1.

\subsection{Electroweak unification}

The singular four-fermion beta-decay interaction does not permit calculations
to higher order.  In a theory where this interaction is the low-energy limit of
massive charged particle ($W$) exchange, however, higher-order calculations
make sense.\cite{GWS} The predicted $W$ and a massive neutral particle, the
$Z$, entailed by the simplest version of the theory, were both discovered in
1983.  The theory also predicted new weak charge-preserving interactions
mediated by $Z$ exchange, first seen in neutrino interactions a decade earlier.
The theory has an SU(2) $\times$ U(1) symmetry, broken by the masses of the
$W$, $Z$, and fermions.

The low-energy limits of $W$ and $Z$ exchange are described by
\begin{equation} \label{eqn:lo}
\frac{G_F}{\sqrt{2}} = \frac{g^2}{8M_W^2} ~~~,~~~
\frac{G_F}{\sqrt{2}} \rho = \frac{g^2 +{g'}^2}{8M_Z^2} ~~~,
\end{equation}
where $G_F$ is the Fermi constant, $g = e/\sin \theta$ and $g' = e/\cos \theta$
are SU(2) and U(1) coupling constants, $e$ is the proton charge, and $\theta$
is the weak mixing angle. The parameter $\rho$, which can arise from effects of
quark loops on $W$ and $Z$ self-energies, is dominated by the top:\cite{Tini}
\begin{equation} \label{eqn:rho}
\rho \simeq 1 + \frac{3G_F m_t^2}{8 \pi^2 \sqrt{2}} ~~~,
\end{equation}
Consequently, if we define $\theta$ by means of the precise measurement
at LEP of $M_Z$,
\begin{equation}
M_Z^2 = \frac{\pi \alpha}{\sqrt{2} G_F \rho \sin^2 \theta \cos^2 \theta}
{}~~~,
\end{equation}
then $\theta$ will depend on $m_t$, and so will
\begin{equation}
M_W^2 = \frac{\pi \alpha}{\sqrt{2} G_F \sin^2 \theta}~~~.
\end{equation}

One must remember that the electric charge becomes stronger at the short
distances characterizing $W$ and $Z$ exchanges as a consequence of vacuum
polarization effects in calculating $M_W$ and $M_Z$ from these expressions.

\subsection{The Higgs boson}

The replacement of the Fermi interaction by massive $W$ exchange does not
cure all the problems of the electroweak theory.  At high energies, $W^+
W^-$ scattering would violate unitarity (i.e., probability conservation)
unless a spinless neutral boson (the {\it Higgs boson}) existed below about
1 TeV.\cite{LQT}  The search for this boson is one motivation for
multi-TeV hadron colliders.  Searches in $e^+ e^- \to {\rm Higgs~boson}
+ \ldots$ at LEP have set a lower limit of about 60 GeV on $M_H$.\cite{SO}
(Here and elsewhere $c = \hbar = 1$.)

The Higgs boson affects the parameter $\rho$ through loop diagrams contributing
to $W$ and $Z$ self-energies.  It is convenient to express contributions to
$\rho$ in terms of deviations of the top quark and Higgs boson masses from
nominal values.  For $m_t = 175$ GeV, $M_H = 300$ GeV, the measured value of
$M_Z$ leads to a nominal expected value of $\sin^2 \theta_0 \equiv x_0 =
0.2320.$  Defining a parameter $T$ of order 1 by $\Delta \rho \equiv \alpha T$,
we find
\begin{equation}
T \simeq \frac{3}{16 \pi \sin^2 \theta} \left[ \frac{m_t^2 - (175
{}~{\rm GeV})^2}{M_W^2} \right] - \frac{3}{8 \pi \cos^2 \theta}
\ln \frac{M_H}{300~{\rm GeV}} ~~~.
\end{equation}
This expression is quadratic in $m_t$, but only logarithmic in $M_H$.  The weak
mixing angle $\theta$, the $W$ mass, and other electroweak observables
depend on $m_t$ and $M_W$.

\subsection{$S$ and $T$ parameters}

The weak charge-changing and neutral-current interactions are probed under a
number of different conditions, corresponding to different values of momentum
transfer.  For example, muon decay occurs at momentum transfers small with
respect to $M_W$, while the decay of a $Z$ into fermion-antifermion pairs
imparts a momentum of nearly $M_Z/2$ to each member of the pair. Although
coupling constants and masses on the right-hand sides of (\ref{eqn:lo}) vary
fairly rapidly with momentum transfer, their quotients vary much less rapidly,
as first pointed out by Veltman\cite{Tini}. Small ``oblique''
corrections,\cite{PT,JRRMP} logarithmic in $m_t$ and $M_H$, arise from
contributions of new particles to the photon, $W$, and $Z$ propagators. Other
(smaller) ``direct'' radiative corrections are important in calcuating actual
values of observables.

We may then replace (\ref{eqn:lo}) by
\begin{equation}
\frac{G_F}{\sqrt{2}} = \frac{g^2}{8 M_W^2} \left( 1 + \frac{\alpha S_W}{4
\sin^2 \theta} \right)~~~,~~~
\frac{G_F \rho}{\sqrt{2}} = \frac{g^2 + {g'}^2}{8M_Z^2} \left( 1 + \frac{\alpha
S_Z}{4 \sin^2 \theta \cos^2 \theta} \right)~~~,
\end{equation}
where $S_W$ and $S_Z$ are coefficients representing variation with momentum
transfer. Together with $T$, they express a wide variety of electroweak
observables in terms of quantities sensitive to new physics.

Expressing the ``new physics'' effects in terms of deviations from nominal
values of top quark and Higgs boson masses, we have the expression for $T$
written above, while contributions of Higgs bosons and of possible new fermions
$U$ and $D$ with electromagnetic charges $Q_U$ and $Q_D$ to $S_W$ and $S_Z$
are\cite{KL}
\begin{equation} \label{eqn:sz}
S_Z = \frac{1}{6 \pi} \left [
\ln \frac{M_H}{300~\g} + \sum N_C \left ( 1 - 4 \overline Q \ln
\frac{m_U}{m_D} \right ) \right ] ~~~,
\end{equation}
\begin{equation} \label{eqn:sw}
S_W = \frac{1}{6 \pi} \left [
\ln \frac{M_H}{300 ~\g} + \sum N_C \left ( 1 - 4 Q_D \ln \frac{m_U}{m_D}
\right ) \right ]~~.
\end{equation}
The expressions for $S_W$ and $S_Z$ are written for doublets of fermions with
$N_C$ colors and $m_U \geq m_D \gg m_Z$, while $\overline Q \equiv (Q_U + Q_D )
/2$. The sums are taken over all doublets of new fermions. In the limit $m_U =
m_D$, one has equal contributions to $S_W$ and $S_Z$. For a single Higgs boson
and a single heavy top quark, Eqs.~(\ref{eqn:sz}) and (\ref{eqn:sw}) become
\begin{equation}
S_Z = \frac{1}{6 \pi} \left [ \ln \frac{M_H}{300~\g} - 2 \ln
\frac{m_t}{175~\g} \right ] ~~;~~~
S_W = \frac{1}{6 \pi} \left [ \ln \frac{M_H}{300~\g} + 4 \ln
\frac{m_t}{175~\g} \right ] ~~.
\end{equation}
We shall use these expressions, together with previous ones, to express all
electroweak observables as functions of $m_t$ and $M_H$.

\subsection{Electroweak experiments}

Recent direct $W$ mass measurements, in GeV, include
$79.92 \pm 0.39$ \cite{OldCDFW},
$80.35 \pm 0.37$ \cite{UA2W},
$80.37 \pm 0.23$ \cite{NewCDFW},
$79.86 \pm 0.26$ \cite{D0W},
with average $80.23 \pm 0.18$ \cite{Wavg}.  Recent data\cite{CCFR,CDHS,CHARM}
on the ratio $R_\nu \equiv \sigma(\nu N \to \nu + \ldots)/\sigma(\nu N \to
\mu^- + \ldots)$ lead to information on $\rho^2$ times a function of $\sin^2
\theta$ roughly equivalent to the constraint $M_W = 80.27 \pm 0.26$ GeV.
Measured $Z$ parameters \cite{LEP} include
$M_Z = 91.1888 \pm 0.0044~{\rm GeV}$,
$\Gamma_Z = 2.4974 \pm 0.0038~{\rm GeV}$,
$\sigma_h^0 = 41.49 \pm 0.12$ nb (the hadron production cross section), and
$R_\ell \equiv \Gamma_{\rm hadrons}/\Gamma_{\rm leptons} = 20.795 \pm 0.040$,
which may be combined to obtain the $Z$ leptonic width $\Gamma_{\ell\ell}(Z) =
83.96 \pm 0.18$ MeV.  Leptonic asymmetries include the forward-backward
asymmetry parameter $A_{FB}^{\ell} = 0.0170 \pm 0.0016$, leading to a value
$\sin^2 \theta_\ell \equiv \sin^2 \theta_{\rm eff}  = 0.23107 \pm 0.0090$, and
independent determinations of $\sin^2 \theta_{\rm eff} = (1/4)(1 -
[g_V^{\ell}/g_A^{\ell}])$ from the parameters $A_\tau \to \sin^2 \theta =
0.2320 \pm 0.0013$, $A_e \to \sin^2 \theta = 0.2330 \pm 0.0014$. The last three
values may be combined to yield $\sin^2 \theta = 0.2317 \pm 0.0007$. We do not
use asymmetries as measured in decays of $Z$ to $b \bar b$ (which may reflect
additional new-physics effects, to $c \bar c$ (which are of limited weight
because of large errors), or to light quarks (for which interpretations are
more model-dependent).  This last result differs by about two standard
deviations from that based on the left-right asymmetry parameter $A_{LR}$
measured with polarized electrons at SLC \cite{SLC}: $\sin^2 \theta = 0.2294
\pm 0.0010$.

Parity violation in atoms, stemming from the interference of $Z$ and photon
exchanges between the electrons and the nucleus, provides further information
on electroweak couplings.  The most precise constraint at present arises from
the measurement of the {\it weak charge} (the coherent vector coupling of the
$Z$ to the nucleus), $Q_W = \rho(Z - N - 4 Z \sin^2 \theta)$, in atomic cesium
\cite{CW}, with the result $Q_W({\rm Cs}) = -71.04 \pm 1.58 \pm 0.88$.  The
first error is experimental, while the second is theoretical \cite{Csrc}.  The
prediction \cite{MR} $Q_W({\rm Cs}) = -73.20 \pm 0.13$ is insensitive to
standard-model parameters \cite{MR,PGHS}; discrepancies are good indications of
new physics (such as exchange of an extra $Z$ boson).

\subsection{Results of fits to electroweak observables}

We have performed a fit to the electroweak observables listed in Table 1.  The
``nominal'' values (including \cite{DKS} $\sin^2 \theta_{\rm eff} = 0.2320$)
are calculated for $m_t = 175$ GeV and $M_H = 300$ GeV.  We use $\Gamma_{\ell
\ell}(Z)$, even though it is a derived quantity, because it has little
correlation with other variables in our fit.  It is mainly sensitive to the
axial-vector coupling $g_A^\ell$, while asymmetries are mainly sensitive to
$g_V^\ell$.  We also omit the total width $\Gamma_{\rm tot}(Z)$ from the fit,
since it is highly correlated with  $\Gamma_{\ell \ell}(Z)$ and mainly provides
information on the value of the strong fine-structure constant $\alpha_s$.
With $\alpha_s = 0.12 \pm 0.01$, the observed total $Z$ width is consistent
with predictions.  The partial width $\Gamma(Z \to b \bar b)$ will be treated
separately below.

Each observable in Table 1 specifies a band in the $S - T$ plane with different
slope, as seen from the ratios of coefficients of $S$ and $T$.  Parity
violation in atomic cesium is sensitive almost entirely to $S$.\cite{MR,PGHS}
The impact of $\sin^2 \theta_{\rm eff}$ determinations on $S$ is considerable.
The leptonic width of the $Z$ is sensitive primarily to $T$. The $W$ mass
specifies a band of intermediate slope in the $S-T$ plane; here we assume $S_W
= S_Z$.

\begin{table}
\begin{center}
\caption{Electroweak observables described in fit}
\medskip
\begin{tabular}{c c c c c c} \hline
Quantity        &   Experimental   &   Nominal    &  Experiment/  &  $S$   &
   $T$   \\
                &      value       &    value     &   Nominal     & Coeff. &
Coeff.   \\ \hline
$Q_W$ (Cs)      & $-71.0 \pm 1.8^{~a)} $  &   $ -73.2^{~b)}$
   & $0.970 \pm 0.025$ &  $-0.80$  & $-0.005$ \\
$M_W$ (GeV)     & $80.24 \pm 0.15^{~c)}$  & $80.320^{~d)}$
   & $0.999 \pm 0.002$ &  $-0.29$  &   0.45   \\
$\Gamma_{\ell\ell}(Z)$ (MeV) & $83.96 \pm 0.18^{~e)}$ & $83.90^{~f)}$
   & $1.001 \pm 0.002$ &  $-0.18$  &   0.78   \\
$\sin^2 \theta_{\rm eff}$ & $0.2317 \pm 0.0007^{~f)}$ & $0.2320$
   & $0.999 \pm 0.003$ &  0.0036   & $-0.0026$ \\
$\sin^2 \theta_{\rm eff}$ & $0.2294 \pm 0.0010^{~g)}$ & $0.2320$
   & $0.989 \pm 0.004$ &  0.0036   & $-0.0026$ \\ \hline
\end{tabular}
\end{center}
\leftline{$^{a)}$ {\small Weak charge in cesium}\cite{CW};
{}~~$^{b)}$ {\small Calculation~\cite{MR} incorporating
atomic physics corrections}~\cite{Csrc}}
\leftline{$^{c)}$ {\small Average of direct measurements~\cite{Wavg}
and indirect information}}
\leftline{{\small \quad from neutral/charged current ratio in
deep inelastic neutrino scattering}\cite{CCFR,CDHS,CHARM}}
\leftline{$^{d)}$ {\small Including perturbative QCD corrections}~\cite{DKS};
{}~~$^{e)}$ {\small LEP average as of July, 1994}\cite{LEP}}
\leftline{$^{f)}$ {\small From asymmetries at LEP}\cite{LEP};
{}~~$^{g)}$ {\small From left-right asymmetry in annihilations at
SLC}\cite{SLC}}
\end{table}

\begin{figure}
\centerline{\epsfysize = 3in \epsffile{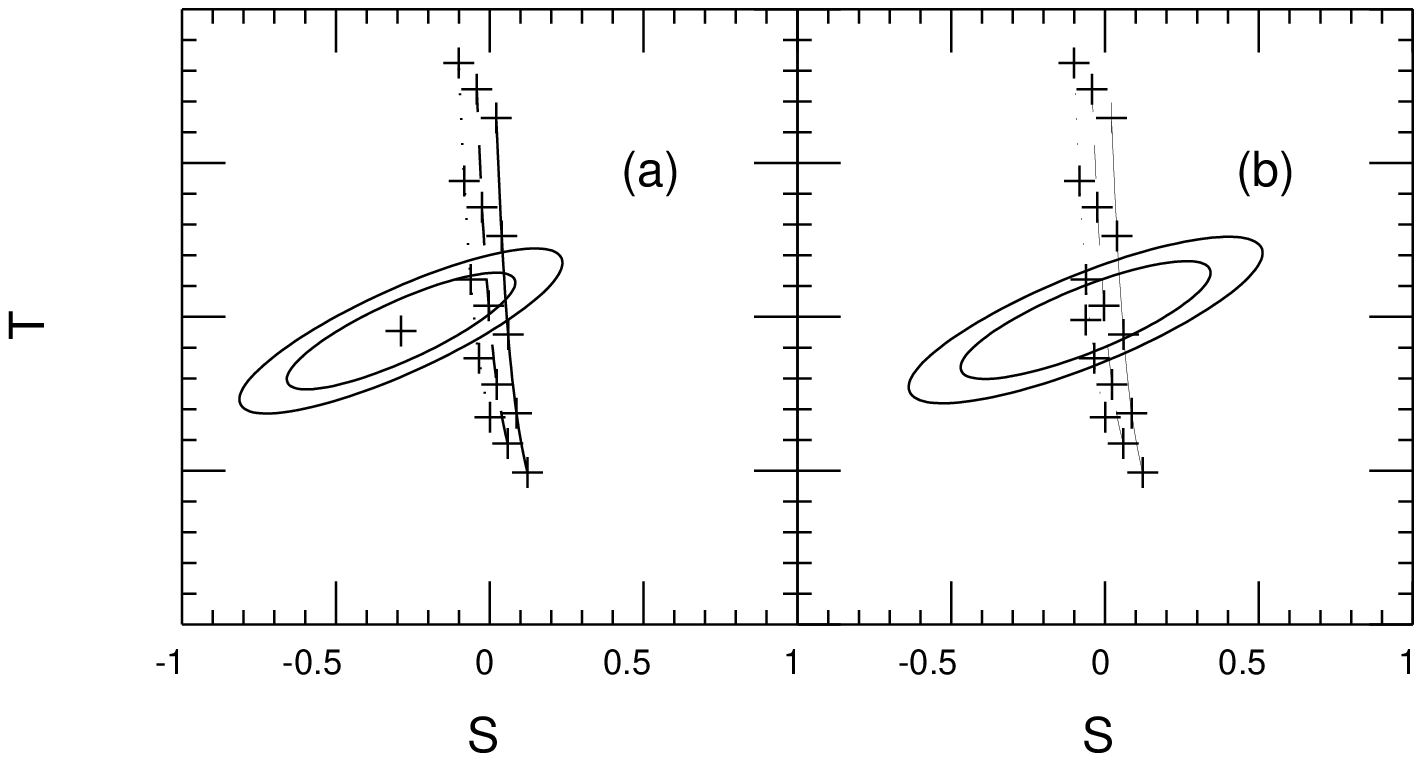}}
\caption{Allowed ranges of $S$ and $T$ at 68\% (inner ellipses) and 90\%
(outer ellipses) confidence levels.  Dotted, dashed, and solid lines correspond
to standard model predictions for $M_H = 100$, 300, 1000 GeV.  Tick marks, from
bottom to top, denote predictions for $m_t = 100$, 140, 180, 220, and 260 GeV.
(a)  Based on data in Table 1; (b) based on data in Table 1 aside from last
row.}
\end{figure}

The resulting constraints on $S$ and $T$ are shown in Figs.~2, both with (a)
and without (b) the SLC data.  Conclusions about the allowed range of $S$ (and
hence about the number of new fermions or other new particles allowed in the
theory) are sensitive to the precise value of $\sin^2 \theta_{\rm eff}$, though
models with large numbers of new fermions (leading to large positive values of
$S$) \cite{PT} are excluded.

We have evaluated the overall quality of a fit to the standard model of the
data in Table 1, when the constraint $m_t = 174 \pm 17$ GeV is added.\cite{DPF}
The result (see also Fig.~5 of Ref.~1) is a minimum $\chi^2 = 8.3$ for
5 degrees of freedom (d.o.f.) at $M_H \approx 260$ GeV.  When the SLC data
on $\sin^2 \theta$ are omitted, the minimum $\chi^2$ falls to 2.4 at $M_H
\approx 700$ GeV for 4 d.o.f.  The error in $M_H$ for $\Delta \chi^2 = 1$
corresponds to about a factor of three in either case.  Conclusions about the
Higgs boson mass clearly are premature!

\subsection{The decay $Z \to b \bar b$}

The measured ratio\cite{LEPHF}$R_b \equiv \Gamma(Z \to b \bar b)/\Gamma(Z \to
{\rm hadrons})$ lies about $2 \sigma$ above the standard model
prediction.\cite{AKG}  If this discrepancy becomes more significant, there
are many possibilities for explaining it within the context of new
physics.\cite{Chiv}  Two-Higgs-doublet models\cite{AKG} cannot reduce the
discrepancy much without running afoul of other constraints (such as the
observed rate\cite{expen} for $b \to s \gamma$).

\section{The CKM Matrix}

The unitary Cabibbo-Kobayashi-Maskawa\cite{cab,KM} (CKM) matrix describes the
weak charge-changing transitions among quarks.  Our present understanding of CP
violation links the observed effect in the neutral kaon system to a phase in
this matrix, whose parameters need to be specified as precisely as possible
in order to test the theory.  Moreover, theories of quark masses necessarily
predict the CKM elements,\cite{JRCKM} and may thereby be tested.

\subsection{Definitions and magnitudes}

We use a convenient parametrization\cite{wp} of the matrix:

\begin{equation}
V = \left ( \begin{array}{c c c}
V_{ud} & V_{us} & V_{ub} \\
V_{cd} & V_{cs} & V_{cb} \\
V_{td} & V_{ts} & V_{tb}
\end{array} \right )
\approx \left [ \matrix{1 - \lambda^2 /2 & \lambda & A \lambda^3 ( \rho -
i \eta ) \cr
- \lambda & 1 - \lambda^2 /2 & A \lambda^2 \cr
A \lambda^3 ( 1 - \rho - i \eta ) & - A \lambda^2 & 1 \cr } \right ]~~~~~ .
\end{equation}

The four parameters are measured as follows:

1.  The parameter $\lambda$ is measured by a comparison of strange particle
decays with muon decay and nuclear beta decay, leading to $\lambda \approx \sin
\theta \approx 0.22$, where $\theta$ is the Cabibbo \cite{cab} angle.

2.  The dominant decays of $b$-flavored hadrons occur via the element $V_{cb}
= A \lambda^2$.  The lifetimes of these hadrons and their semileptonic
branching ratios then lead to an estimate $A = 0.79 \pm 0.06$.

3.  The decays of $b$-flavored hadrons to charmless final states allow one to
measure the magnitude of the element $V_{ub}$ and thus to conclude that
$|V_{ub}/V_{cb}| = 0.08 \pm 0.02$ or $\sqrt{\rho^2 + \eta^2} = 0.36 \pm 0.09$.

4.  The least certain quantity is the phase of $V_{ub}$:  Arg $(V_{ub}^*) =
\arctan(\eta/\rho)$.  Information on this quantity may be obtained by studying
its effect on contributions of higher-order diagrams involving the top quark.

The unitarity of the matrix (up to order $\lambda^3$) is implicit in the
parametrization; in particular, the relation $V_{ub}^* + V_{td} \simeq A
\lambda^3$, when normalized so that its right-hand side is unity, becomes $\rho
+ i \eta  + (1 - \rho - i \eta) = 1$.  We shall be concerned with the allowed
region in the $(\rho,\eta)$ plane.

\subsection{Indirect information}

Two important sources of indirect information on $\rho$ and $\eta$ are $B^0 -
\overline{B}^0$ mixing and CP-violating $K^0 - \overline{K}^0$ mixing.

The presence of ``wrong-sign'' leptons in semileptonic $B$ decays provided the
first evidence for $B^0 - \overline{B}^0$ mixing.\cite{bmix}  Many experiments,
including recent observations of time-dependent oscillations,\cite{newmix} have
confirmed the effect, leading to a mixing amplitude $\Delta m/\Gamma = 0.71 \pm
0.07$.  The dominant contribution to the mixing is expected to arise from
one-loop diagrams (``box graphs'') involving internal $W$ and top quark lines,
leading to $\Delta m \sim f_B^2 m_t^2 |V_{td}|^2$ (times a slowly varying
function of $m_t/M_W$). Here the ``$B$ decay constant,'' $f_B$, describes the
amplitude for finding a $b$ antiquark and a light quark at the same point in a
$B$ meson.  Since $|V_{td}| \sim | 1 - \rho - i \eta|$, the $B^0 -
\overline{B}^0$ mixing amplitude leads to a constraint in the $(\rho,\eta)$
plane consisting of a circular band with center (1,0).  The main contribution
to the width of this band is uncertainty in $f_B$.

A similar set of box diagrams contributes to the parameter $\epsilon$
describing CP-violating $K^0 - \overline{K}^0$ mixing.  The imaginary part of
the mass matrix is proportional to $f_K^2 m_t^2 {\rm Im} (V_{td}^2)$ times a
slowly varying function of $m_t$, with a small correction for the charmed quark
contribution and an overall factor $B_K$ describing the degree to which the box
graphs account for the effect.  Since Im($V_{td}^2) \sim \eta (1 - \rho)$, the
constraint imposed by CP-violating $K^0 - \overline{K}^0$ mixing consists of a
hyperbolic band in the $(\rho,\eta)$ plane with focus at (1,0), whose width is
dominated by uncertainty in the magnitude of $V_{cb}$.\cite{StonePas}

\subsection{Constraints on $\rho$ and $\eta$}

The allowed region\cite{DPF} in $(\rho,\eta)$ is shown in Fig.~3.  Parameters
used, in addition to those mentioned above, include $B_K = 0.8 \pm 0.2$, $f_B =
180 \pm 30$ MeV (in units where $f_\pi = 132$ MeV), $\eta_{\rm QCD} = 0.6 \pm
0.1$ (a correction to the $B - \overline{B}$ mixing diagrams), and $B_B = 1$
for the factor analogous to $B_K$. The center of the allowed region lies around
$\rho \simeq 0,~ \eta \simeq 0.35$.  We may ask how our information may be
improved.

\subsection{Improved tests}

We shall concentrate on a few key aspects of $B$ physics, since extensive
discussions of processes involving rare kaon decays and the CP-violating
parameter $\epsilon'/\epsilon$ appear elsewhere.\cite{DPF,JRCKM,JRCP}

{\it 3.4.1 Decay constant information.} The value of $f_B$ affects the
interpretation of $B - \bar B$ mixing in terms of $|1 - \rho - i \eta|$. One
recent lattice gauge theory calculation obtains\cite{BLS}
$f_B = 187 \pm 10 \pm 34 \pm 15$ MeV,
$f_{B_s} = 207 \pm 9 \pm 34 \pm 22$ MeV,
$f_D = 208 \pm 9 \pm 35 \pm 12$ MeV,
$f_{D_s} = 230 \pm 7 \pm 30 \pm 18$ MeV,
where the first error is statistical, the second is associated with
fitting and lattice size, and the third arises from scaling.  Several
other determinations exist,\cite{lats} some of which obtain values outside
the limits just cited.

\begin{figure}
\centerline{\epsfysize = 4in \epsffile{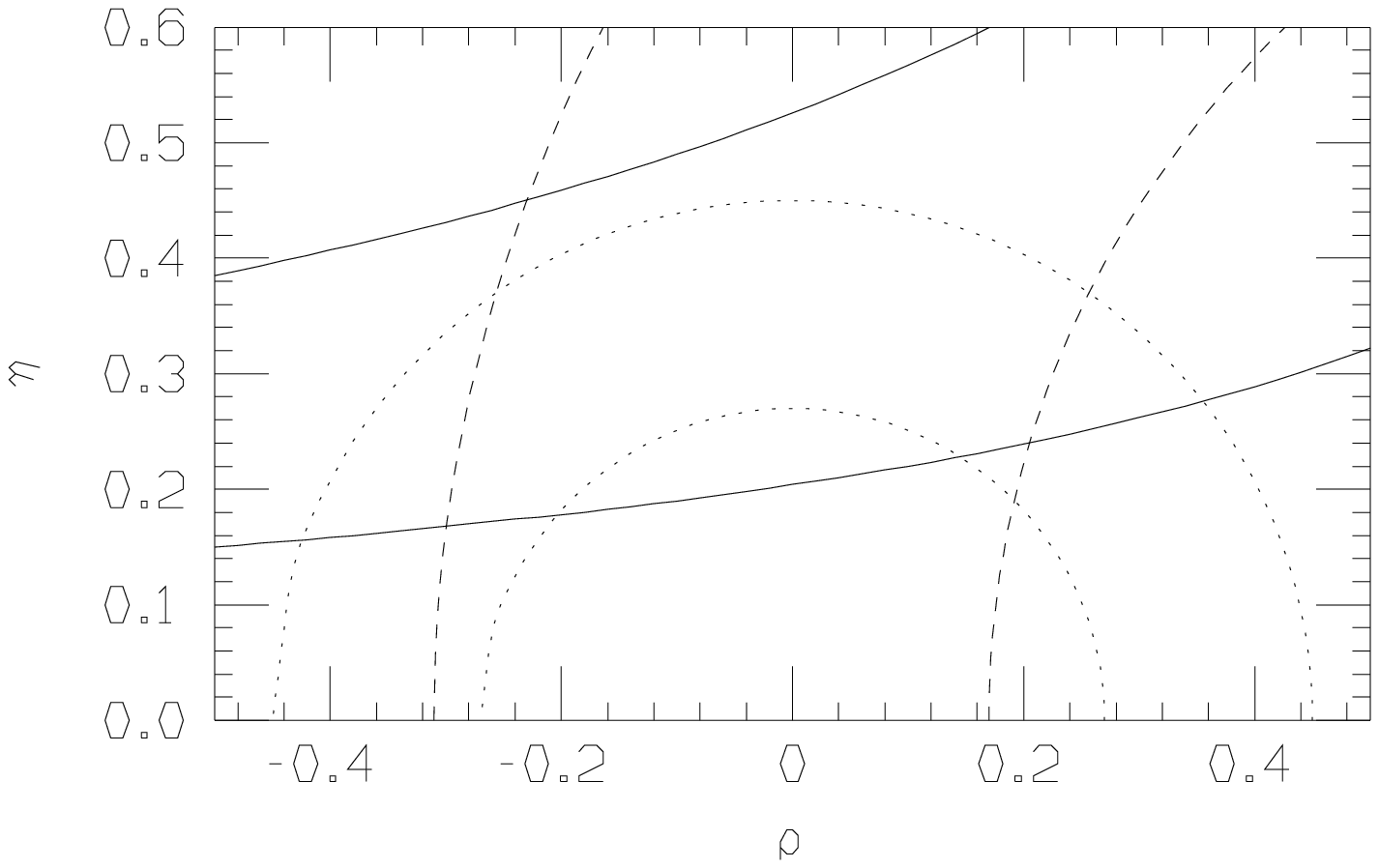}}
\caption{Region in the $(\rho,\eta)$ plane allowed by various constraints.
Dotted semicircles denote central value and $\pm 1 \sigma$ limits implied by
$|V_{ub}/V_{cb}| = 0.08 \pm 0.02$.  Circular arcs with centers at $(\rho,\eta)
= (1,0)$ denote constraints from $B - \overline{B}$ mixing, while hyperbolae
describe region bounded by constraints from CP-violating $K - \overline{K}$
mixing.}
\end{figure}

The WA75 Collaboration\cite{WA75} reports a handful of $D_s \to \mu \nu$
candidates, leading to $f_{D_s} = 232 \pm 69$ MeV, while the CLEO
Collaboration,\cite{FDSCLEO} based on several dozen such events and an average
by Muheim and Stone,\cite{MS} obtains $f_{D_s} = 315 \pm 46$ MeV.  We average
these two values to obtain $f_{D_s} = 289 \pm 38$ MeV. The BES group also has a
few candidates for $D_s \to \mu \nu$ or $D_s \to \tau \nu$;\cite{BES} the E653
Collaboration has candidates for $D_s \to \mu \nu$ for which results are
expected soon.\cite{E653}

The lattice determination noted above obtains $f_D/f_{D_s} \simeq 0.9$,
while an argument based on the quark model and the observed hyperfine
splittings between singlet and triplet charmed nonstrange and strange mesons
\cite{AM} suggests $f_D/f_{D_s} \simeq 0.8$.  Using this last ratio, we
obtain $f_D = 231 \pm 31$ MeV, not far below the present upper limit
(90\% c.l.) of 290 MeV obtained by the Mark III Collaboration\cite{MKIII}
on the basis of their search for the decay $D^+ \to \mu^+ \nu$.  The BES
Collaboration should be able to observe this process in events obtained in
the reaction $e^+ e^- \to \psi(3770) \to D^+ D^-$.

Heavy meson decay constants are related to the square of the quark-antiquark
wave function at the origin,\cite{ES} which can be estimated\cite{AM} with the
help of spin-dependent isospin mass splittings in $K$, $D$, and $B$ systems.
The end result of this exercise\cite{DPF,AM} is the result $f_B = 180$ MeV
employed above.  The error is based on the spread in various theoretical
estimates.

{\it 3.4.2 $B$ Decays to CP non-eigenstates: pair of light pseudoscalars.}
A difference between the rates for a process and its charge-conjugate, such as
$B^+ \to \pi^+ K^0$ and $B^- \to \pi^- K^0$, signifies CP violation.  Under
charge conjugation, weak phases change sign, but strong phases do not.  In
order for a rate difference to appear, there must be both a weak phase
difference and a strong phase difference in the channels with isospins
$I = 1/2$ and 3/2.  Recently it has been shown that one may be able to
measure weak phases via the rates for $B$ decays to pairs of light pseudoscalar
mesons {\it without} having any strong phase differences.\cite{BPP}  The
presence of electroweak penguins\cite{DH} is one possible obstacle to this
program, which is under further investigation.

{\it 3.4.3 Decays of neutral $B$ mesons to CP eigenstates: $\pi - B$
correlations.}  Although produced initially as flavor eigenstates, neutral $B$
mesons can undergo $B^0 - \overline{B}^0$ mixing, leading to time-dependent
asymmetries in decays to CP eigenstates like $J/\psi K_S$.  Time-integrated
decays also can display rate asymmetries, whose interpretation is often
independent of final-state effects.  For example, the asymmetry in decays of
$B^0$ or $\overline{B}^0$ to $J/\psi K_S$ is equal to $-[x_d/(1 + x_d^2)]\sin
[{\rm Arg} (V_{td}^*)^2]$, where $x_d = (\Delta m / \Gamma)|_d = 0.70 \pm 0.07$
is the mixing parameter mentioned earlier.  One has to know the flavor of the
neutral $B$ at time of production.  One proposed means for ``tagging'' the
$B$ involves its correlation with charged pions produced nearby in phase
space.\cite{GNR}  The existence of such a correlation is predicted both by
fragmentation and resonance decay pictures.

\section{Intermediate mass scales}

\subsection{Neutrino masses and new mass scales}

The pattern of neutrino masses in Fig.~1 is at least as puzzling as the
large mass of the top quark.  Why are neutrinos so light?

One proposed solution to the apparent suppression of the solar neutrino
flux, especially for neutrinos with energies of a few MeV,\cite{SNU} is based
on matter-induced oscillations between electron neutrinos and some other
species.\cite{MSW} An acceptable solution involves a muon neutrino mass of
several millielectron volts and a much lighter electron neutrino.  A
large Majorana mass\cite{seesaw} of $10^9 - 10^{12}$ GeV for the right-handed
neutrino, combined with a Dirac mass of order 1 GeV linking the right-handed
and left-handed muon neutrino, could give such a mass eigenstate.

If the corresponding tau neutrino Dirac mass is related to the top quark mass
(as it is in some grand unified theories), the $\nu_\tau$ could then have a
mass of a few eV.  It might well mix with the $\nu_\mu$ with an angle $\theta
\ge m_\mu/m_\tau$, leading to a mixing parameter $\sin^2 2 \theta \ge 10^{-2}$.

Present limits\cite{E531} from Fermilab Experiment E531 restrict the muon and
tau neutrinos to have $\Delta m^2 \leq 1 {\rm~eV}^2$ for large $\theta$ and
$\sin^2 2 \theta \leq {\rm~(a~few)} \times 10^{-3}$ for large $\Delta m^2$. A
wider range of parameters are accessible to CHORUS,\cite{CHORUS} an emulsion
experiment, and NOMAD,\cite{NOMAD} a fine-grained detector, both operating at
CERN, which should reach $\sin^2 2 \theta \leq {\rm~(a~few)} \times 10^{-4}$
for large $\Delta m^2$.  New short- and long-baseline $\nu_\mu \leftrightarrow
\nu_\tau$ oscillation experiments\cite{E803,P822} are also envisioned for
Fermilab, while experiments on atmospheric neutrinos\cite{KAM} and low-energy
reactor neutrinos\cite{LSND} have recently presented tantalizing hints of
oscillation phenomena.\cite{FH}  A new solar-neutrino experiment, sensitive
mainly to high-energy electron neutrinos but eventually to other species as
well through neutral-current interactions, will begin operating in a couple of
years.\cite{SNO}

\subsection{Electroweak-strong unification}

The U(1), SU(2), and SU(3) couplings of the electroweak and strong interactions
approach one another at high energies,\cite{GQW} but do not really meet at a
single point, in contrast to the predictions of the simplest SU(5) grand
unification scheme.\cite{GG}  One cure\cite{Amaldi} for this ``astigmatism'' is
provided by the supersymmetric extension of SU(5), in which the inclusion of
superpartners below about 1 TeV modifies the slopes in the relations between
$\alpha_i^{-1}~(i=1,2,3)$ and $\ln M^2$ such that the couplings meet slightly
above $10^{16}$ GeV at $\alpha_i^{-1} \approx 26$. Another possibility,
however, is to embed SU(5) in an SO(10) model,\cite{SOten} with SO(10) breaking
to SU(3)$_{\rm color} \times$ SU(2)$_{\rm R} \times$ SU(2)$_{\rm L} \times$
U(1)$_{\rm B-L}$ around $4 \times 10^{17}$ GeV and SU(2)$_{\rm R} \times$
U(1)$_{\rm B-L}$ breaking to U(1)${\rm Y}$ at an intermediate mass scale of
about $10^{10}$ GeV.  (An early analysis\cite{RR} indicated that this
intermediate scale could lie anywhere between $10^9$ and about $10^{12}$ GeV.)
These possibilities are compared in Fig.~9 of Ref.~1.

\subsection{Baryogenesis}

In 1967 Sakharov\cite{Sakh} identified three key elements of any theory seeking
to explain the observed baryon abundance of the Universe, $n_B/n_\gamma \simeq
4 \times 10^{-9}$:  (1)  C and CP violation; (2) baryon number violation, and
(3) a period during which the Universe was out of thermal equilibrium.  A toy
model is provided by the decay of SU(5) ``$X$'' bosons, which couple both to
pairs of quarks and to antiquark-lepton pairs.  The total rates for $X$ and
$\overline{X}$ decays must be the same as a consequence of CPT invariance, but
the branching ratios $B(X \to uu)$ and $B(X \to e^+ \bar d)$ can differ from
$B(\overline{X} \to \bar u \bar u)$ and $B(\overline{X} \to e^- d)$ as a result
of CP violation, leading to a baryon asymmetry once the Universe cools below
the energy where these decay processes are balanced by the reverse processes.
This example conserves $B - L$, where $B$ is baryon number (1/3 for quarks) and
$L$ is lepton number (1 for electrons).

It was pointed out by 't Hooft \cite{tH} that the electroweak theory contains
an anomaly as a result of nonperturbative effects which conserve $B - L$ but
violate $B + L$.  If a theory leads to $B - L = 0$ but $B + L \ne 0$ at some
primordial temperature $T$, the anomaly can wipe out any $B+L$ as $T$ sinks
below the electroweak scale \cite{KRS}. Thus, the toy model mentioned above and
many others are unsuitable in practice.

One proposed solution is the generation of nonzero $B - L$ at a high
temperature, e.g., through the generation of nonzero lepton number $L$, which
is then reprocessed into nonzero baryon number by the `t Hooft anomaly
mechanism \cite{Yana}. The existence of a baryon asymmetry, when combined with
information on neutrinos, could provide a window to a new scale of particle
physics. Large Majorana masses acquired by right-handed neutrinos would change
lepton number by two units and thus would be ideal for generating a lepton
asymmetry if Sakharov's other two conditions are met.

The question of baryogenesis is thus shifted onto the leptons:  Do neutrinos
indeed have masses?  If so, what is their ``CKM matrix''?  Do the properties of
heavy Majorana right-handed neutrinos allow any new and interesting natural
mechanisms for violating CP at the same scale where lepton number is violated?
Majorana masses for right-handed neutrinos naturally violate left-right
symmetry and could be closely connected with the violation of $P$ and $C$ in
the weak interactions \cite{BKCP}. An open question in this scenario, besides
the form of CP violation at the lepton-number-violating scale, is the manner in
which CP violation is communicated to the lower mass scale at which we see CKM
phases.

\subsection{The strong CP problem}

The Lagrangian density of QCD in principle can contain a term $(g_3^2 \bar
\theta /32 \pi^2) F^a_{\mu \nu} \tilde{F}^{\mu \nu a}$, where $\bar \theta
\equiv \theta + {\rm~Arg~det~}M$, with $M$ the quark mass matrix and $\theta$
an angle characterizing the vacuum structure.\cite{RP}  The neutron electric
dipole moment receives a contribution of order $d_n \simeq 10^{-16}\bar \theta
{}~e \cdot$ cm, implying $\bar \theta \leq 10^{-9}$.  How can we understand the
small value of $\bar \theta$ without fine-tuning?

Proposed solutions include (1) an unconventional and still controversial
interpretation of the vacuum\cite{RGS} as composed of an incoherent mixture
of $\theta$ and $-\theta$, in the sense of a density matrix; (2) vanishing
of one of the quark masses (e.g., $m_u$), which probably bends chiral
symmetry beyond plausible limits;\cite{HL} (3) the promotion\cite{PQ} of $\bar
\theta$ to the status of a dynamical variable relaxing to zero, implying the
existence\cite{ax} of a light Nambu-Goldstone boson, the {\it axion}.  Searches
for this particle by means of RF cavities\cite{PS} turn out to be uniquely
sensitive to the scale of symmetry breaking in the range we have been
discussing, $10^9 - 10^{12}$ GeV!

\section{Electroweak symmetry breaking}

A key question facing the standard model of electroweak interactions is the
mechanism for breaking SU(2) $\times$ U(1).  We discuss two popular
alternatives; Nature may turn out to be cleverer than either.

\subsection{Fundamental Higgs boson(s)}

If there really exists a relatively light fundamental Higgs boson in the
context of a grand unified theory, one has to protect its mass from
large corrections.  Supersymmetry is the popular means for doing so.  Then
one expects a richer neutral Higgs structure, charged scalar bosons, and
superpartners, all below about 1 TeV.

\subsection{Strongly interacting Higgs sector}

The scattering of longitudinally polarized $W$ and $Z$ bosons violates
unitarity above a TeV or two if there does not exist a Higgs boson below this
energy \cite{LQT}.  The behavior is similar to that of pion-pion scattering in
the non-linear sigma model above a few hundred MeV, where we wouldn't trust the
model. Similarly, we mistrust the present version of electroweak theory above a
TeV.  If the theory has a strongly interacting sector, its $I = J = 0$ boson
(like the $\sigma$ of QCD) may be its least interesting and most elusive
feature. The rich spectrum of resonances in QCD is now understood in terms of
the interactions of quarks and gluons.  We hope that the exploration of the
potentially rich TeV physics of the electroweak sector can proceed through the
use of both hadron and electron-positron colliders.

\section{Summary}

Increasingly precise electroweak and top quark measurements show promise of
shedding indirect light on the Higgs sector, when present accuracies are
improved by about a factor of three.  Electron-positron collisions at
LEP and SLC, hadron collisions at Fermilab, and even table-top experiments
on atomic parity violation all will play key roles.

The Cabibbo-Kobayashi-Maskawa (CKM) matrix is the leading candidate for
the source of the observed CP violation in the neutral kaon system, through
phases which require the existence of the third family of quarks.  The
observation of events consistent with a top quark has given encouragement
to this scheme, but further confirmation is needed, through the study of
CP-violating and other rare kaon decays and through systematic exploration
of the properties of hadrons containing $b$ quarks.

Several prospective windows exist on the ``intermediate'' mass scale of $10^9 -
10^{12}$ GeV, including the study of neutrino masses, partial unification of
interactions, baryogenesis, and axions.  Thus, there appears to be at least a
lamppost, if not an oasis, in the grand desert between the TeV scale and the
unification scale.

Candidates for understanding the Higgs sector include a theory of fundamental
scalars with masses protected by supersymmetry and a composite-Higgs theory
which could, in principle, be merely the first hints of compositeness on a
deeper level.  Perhaps even quarks and leptons are composite.  I, for one would
not be distressed if we were simply uncovering one more layer of the onion in
our journey toward the Planck scale.

\section{Acknowledgements}

I am grateful to Jim Amundson, Aaron Grant, Michael Gronau, Oscar Hern\'andez,
Nahmin Horowitz, Mike Kelly, David London, Sheldon Stone, Tatsu Takeuchi, and
Mihir Worah for fruitful collaborations on some of the topics mentioned here,
to them and to I. Dunietz, M. Goodman, and G. Zapalac for discussions, to the
Fermilab Theory Group for hospitality during preparation of this lecture, and
to our hosts in Warsaw for providing an enjoyable and informative workshop and
a chance to get acquainted with their fine city. This work was supported in
part by the United States Department of Energy under Grant No. DE AC02
90ER40560.

\def \ap#1#2#3{{\it Ann. Phys. (N.Y.)} {\bf#1} (#3) #2}
\def \apny#1#2#3{{\it Ann.~Phys.~(N.Y.)} {\bf#1} (#3) #2}
\def \app#1#2#3{{\it Acta Physica Polonica} {\bf#1} (#3) #2}
\def \arnps#1#2#3{{\it Ann. Rev. Nucl. Part. Sci.} {\bf#1} (#3) #2}
\def \arns#1#2#3{{\it Ann. Rev. Nucl. Sci.} {\bf#1} (#3) #2}
\def \ba88{{\it Particles and Fields 3} (Proceedings of the 1988 Banff Summer
Institute on Particles and Fields), edited by A. N. Kamal and F. C. Khanna
(World Scientific, Singapore, 1989)}
\def \baphs#1#2#3{{\it Bull. Am. Phys. Soc.} {\bf#1} (#3) #2}
\def \be87{{\it Proceedings of the Workshop on High Sensitivity Beauty
Physics at Fermilab,} Fermilab, Nov. 11-14, 1987, edited by A. J. Slaughter,
N. Lockyer, and M. Schmidt (Fermilab, Batavia, IL, 1988)}
\def \cn{Collaboration}
\def \cp89{{\it CP Violation,} edited by C. Jarlskog (World Scientific,
Singapore, 1989)}
\def \dpf91{{\it The Vancouver Meeting - Particles and Fields '91}
(Division of Particles and Fields Meeting, American Physical Society,
Vancouver, Canada, Aug.~18-22, 1991), ed. by D. Axen, D. Bryman, and M. Comyn
(World Scientific, Singapore, 1992)}
\def \dpff{{\it The Fermilab Meeting - DPF 92} (Division of Particles and
Fields
Meeting, American Physical Society, Fermilab, 10 -- 14 November, 1992), ed. by
C. H. Albright \ite~(World Scientific, Singapore, 1993)}
\def \efi{Enrico Fermi Institute Report No.~}
\def \hb87{{\it Proceeding of the 1987 International Symposium on Lepton and
Photon Interactions at High Energies,} Hamburg, 1987, ed. by W. Bartel
and R. R\"uckl (Nucl.~Phys.~B, Proc. Suppl., vol. 3) (North-Holland,
Amsterdam, 1988)}
\def \ib{{\it ibid.}~}
\def \ibj#1#2#3{{\it ibid.} {\bf#1} (#3) #2}
\def \ijmpa#1#2#3{{\it Int.~J. Mod.~Phys.}~A {\bf#1} (#3) #2}
\def \ite{{\it et al.}}
\def \jpb#1#2#3{{\it J. Phys.} B {\bf#1} (#3) #2}
\def \jpg#1#2#3{{\it J. Phys.} G {\bf#1} (#3) #2}
\def \kdvs#1#2#3{{\it Kong.~Danske Vid.~Selsk., Matt-fys.~Medd.} {\bf #1}
(#3) No #2}
\def \ky85{{\it Proceedings of the International Symposium on Lepton and
Photon Interactions at High Energy,} Kyoto, Aug.~19-24, 1985, edited by M.
Konuma and K. Takahashi (Kyoto Univ., Kyoto, 1985)}
\def \lat90{{\it Results and Perspectives in Particle Physics} (Proceedings of
Les Rencontres de Physique de la Vallee d'Aoste [4th], La Thuile, Italy, Mar.
18-24, 1990), edited by M. Greco (Editions Fronti\`eres, Gif-Sur-Yvette,
France,
1991)}
\def \lg91{International Symposium on Lepton and Photon Interactions, Geneva,
Switzerland, July, 1991}
\def \lkl87{{\it Selected Topics in Electroweak Interactions} (Proceedings of
the Second Lake Louise Institute on New Frontiers in Particle Physics, 15 --
21 February, 1987), edited by J. M. Cameron \ite~(World Scientific, Singapore,
1987)}
\def \mpla #1#2#3{{\it Mod. Phys. Lett.} A {\bf#1} (#3) #2}
\def \nc#1#2#3{{\it Nuovo Cim.} {\bf#1} (#3) #2}
\def \np#1#2#3{{\it Nucl. Phys.} {\bf#1} (#3) #2}
\def \oxf65{{\it Proceedings of the Oxford International Conference on
Elementary Particles} 19/25 Sept.~1965, ed.~by T. R. Walsh (Chilton, Rutherford
High Energy Laboratory, 1966)}
\def \pisma#1#2#3#4{{\it Pis'ma Zh. Eksp. Teor. Fiz.} {\bf#1} (#3) #2 [{\it
JETP Lett.} {\bf#1} (#3) #4]}
\def \pl#1#2#3{{\it Phys. Lett.} {\bf#1} (#3) #2}
\def \plb#1#2#3{{\it Phys. Lett.} B {\bf#1} (#3) #2}
\def \ppnp#1#2#3{{\it Prog. Part. Nucl. Phys.} {\bf#1} (#3) #2}
\def \pr#1#2#3{{\it Phys. Rev.} {\bf#1} (#3) #2}
\def \prd#1#2#3{{\it Phys. Rev.} D {\bf#1} (#3) #2}
\def \prl#1#2#3{{\it Phys. Rev. Lett.} {\bf#1} (#3) #2}
\def \prp#1#2#3{{\it Phys. Rep.} {\bf#1} (#3) #2}
\def \ptp#1#2#3{{\it Prog. Theor. Phys.} {\bf#1} (#3) #2}
\def \rmp#1#2#3{{\it Rev. Mod. Phys.} {\bf#1} (#3) #2}
\def \si90{25th International Conference on High Energy Physics, Singapore,
Aug. 2-8, 1990, Proceedings edited by K. K. Phua and Y. Yamaguchi (World
Scientific, Teaneck, N. J., 1991)}
\def \slac75{{\it Proceedings of the 1975 International Symposium on
Lepton and Photon Interactions at High Energies,} Stanford University, Aug.
21-27, 1975, edited by W. T. Kirk (SLAC, Stanford, CA, 1975)}
\def \slc87{{\it Proceedings of the Salt Lake City Meeting} (Division of
Particles and Fields, American Physical Society, Salt Lake City, Utah, 1987),
ed. by C. DeTar and J. S. Ball (World Scientific, Singapore, 1987)}
\def \smass82{{\it Proceedings of the 1982 DPF Summer Study on Elementary
Particle Physics and Future Facilities}, Snowmass, Colorado, edited by R.
Donaldson, R. Gustafson, and F. Paige (World Scientific, Singapore, 1982)}
\def \smass90{{\it Research Directions for the Decade} (Proceedings of the
1990 DPF Snowmass Workshop), edited by E. L. Berger (World Scientific,
Singapore, 1991)}
\def \smassb{{\it Proceedings of the Workshop on $B$ Physics at Hadron
Accelerators}, Snowmass, Colorado, 21 June -- 2 July 1993, ed.~by P. McBride
and C. S. Mishra, Fermilab report FERMILAB-CONF-93/267 (Fermilab, Batavia, IL,
1993)}
\def \stone{{\it B Decays}, edited by S. Stone (World Scientific, Singapore,
1994)}
\def \tasi90{{\it Testing the Standard Model} (Proceedings of the 1990
Theoretical Advanced Study Institute in Elementary Particle Physics),
edited by M. Cveti\v{c} and P. Langacker (World Scientific, Singapore, 1991)}
\def \ufn#1#2#3#4#5#6{{\it Usp.~Fiz.~Nauk} {\bf#1} (#3) #2 [Sov.~Phys. -
Uspekhi {\bf#4} (#6) #5]}
\def \yaf#1#2#3#4{{\it Yad. Fiz.} {\bf#1} (#3) #2 [Sov. J. Nucl. Phys. {\bf #1}
 (#3) #4]}
\def \zhetf#1#2#3#4#5#6{{\it Zh. Eksp. Teor. Fiz.} {\bf #1} (#3) #2 [Sov.
Phys. - JETP {\bf #4} (#6) #5]}
\def \zhetfl#1#2#3#4{{\it Pis'ma Zh. Eksp. Teor. Fiz.} {\bf #1} (#3) #2 [JETP
Letters {\bf #1} (#3) #4]}
\def \zp#1#2#3{{\it Zeit. Phys.} {\bf#1} (#3) #2}
\def \zpc#1#2#3{{\it Zeit. Phys.} C {\bf#1} (#3) #2}

\end{document}